\documentclass[prb,aps,twocolumn]{revtex4}

\usepackage{epsfig}
\usepackage{amsfonts}
\usepackage{amsmath}

\begin{document}

\title{High-field NMR study of the spin correlations in the spin-cluster mineral Na$_2$Cu$_3$O(SO$_4$)$_3$}

\author{Long Ma$^{1}$}
\email{malong@hmfl.ac.cn}
\author{J. X. Li$^{1}$}
\author{L. S. Ling$^{1}$}
\author{Y. Y. Han$^{1}$}
\author{L. Zhang$^{1}$}
\author{L. Hu$^{1}$}
\email{hulin@hmfl.ac.cn}
\author{W. Tong$^{1}$}
\author{C. Y. Xi$^{1}$}
\author{Li Pi$^{1,2}$}

\affiliation{$^{1}$Anhui Province Key Laboratory of Condensed Matter Physics at Extreme Conditions, High Magnetic Field Laboratory, Chinese Academy of Sciences, Hefei 230031, China\\
$^{2}$ Hefei National Research Center for Physical Sciences at the Microscale, University of Science and Technology of China, Hefei 230026, China
}

\date{\today}

\begin{abstract}
  We report NMR study on the spin correlations in the spin-cluster based mineral Na$_2$Cu$_3$O(SO$_4$)$_3$ with magnetic fields ranged from 1 T to 33 T. The long-range magnetic order is observed from both the sudden spectral broadening at $T_N$ and critical slowing down behavior in the temperature dependence of spin-lattice relaxation rates ($1/T_1(T)$). The hump behavior of $1/T_1(T)$ persists to $\mu_0H=7.25$ T, above which a spin excitation gap is observed from the thermally activated temperature dependence of $1/T_1$. The gap size shows a linear field dependence, whose slope and intercept respectively yield an effective magnetic moment of 2.54 $\mu_B$ and a 0.94 meV spin excitation gap under zero magnetic field. These results indicate the existence of short-range order and prominent easy-plane spin anisotropy, which are important for understanding the spin excitation spectrum in A$_2$Cu$_3$O(SO$_4$)$_3$.

\end{abstract}

\maketitle

Novel quantum phenomena can be realized in either periodic arrays of spin-clusters or lattices of magnetic ions. Recently, the minerals A$_2$Cu$_3$O(SO$_4$)$_3$ (A=K, Na or A$_2$=KNa) attract many research interests in the field of quantum magnetism after their discovery in the natural world\cite{Starova_MM_55_613,Lander_IC_56_2013,Siidra_EJM_29_499}. Every Six CuO$_4$ plaquettes form one Cu$_6$-hexamer, which behaves as the cluster with effective spin $S=1$ at low temperatures\cite{Fujihala_PRL_120_077201}. The strong intra-hexamer exchange strength and eight lowest-lying states of the hexamer are unambiguously determined by inelastic neutron scattering (INS)\cite{Furrer_PRB_98_180410}.

The magnetic interaction between Cu$_6$-hexamers is studied from both experimental and theoretical approach\cite{Fujihala_PRL_120_077201,Furrer_PRB_104_L220401,Nekrasova_PRB_102_184405}. For all the three compounds, the magnetic susceptibilities share a similar temperature dependence\cite{Nekrasova_PRB_102_184405}. They first show Curie-Weiss-like upturn upon cooling, then drop at lower temperatures, leading to a broad maxima around $T=6\sim7$ K. The magnetic ordering at $T_N=2.9\sim4.7$ K is observed in the temperature dependence of specific heat or magnetic susceptibility\cite{Nekrasova_PRB_102_184405}. Below $T_N$, opening of the spin excitation gap is verified by INS measurements\cite{Fujihala_PRL_120_077201,Furrer_PRB_98_180410}. At $T=1.5$ K, the excitation energy from the $S=1$ ground state ranges from 0.6 meV to 1.7 meV for A= Na and K\cite{Furrer_PRB_98_180410}.
This gapped excitation is first proposed as a consequence of the cluster-based Haldane state\cite{Fujihala_PRL_120_077201}. Contrastively, the \emph{ab initio} calculations demonstrate the two-dimensional inter-hexamer interactions in the crystalline $bc$-plane\cite{Nekrasova_PRB_102_184405}. The spin gap is attributed to the magnetic anisotropy of the square lattice, instead. The subsequent INS measurements support the two-dimensional characteristic of the magnetic interactions\cite{Furrer_PRB_104_L220401},while no direct evidence for the magnetic anisotropy is reported up to now.

Studying how the spin system responds to applied magnetic fields is a useful way to explore the magnetic interactions in quantum magnets. For the present A$_2$Cu$_3$O(SO$_4$)$_3$ compound, previous INS studies were performed under a zero magnetic field\cite{Fujihala_PRL_120_077201,Furrer_PRB_98_180410,Furrer_PRB_104_L220401}, thus the field dependence of the spin excitations is still lacked although highly demanded. As a local probe sensitive to low-energy spin excitations, nuclear magnetic resonance (NMR) is very suitable for tracking the evolution of spin excitations under magnetic fields.

In this article, we employ NMR to study the spin excitation evolution in Na$_2$Cu$_3$O(SO$_4$)$_3$ with applied magnetic fields up to 33 T. The magnetic ordering is monitored by the sudden spectral broadening across $T_N$ at the low field side, also the critical slowing down behavior near $T_N$
from the temperature dependence of spin-lattice relaxation rates(SLRRs). For $H_c<\mu_0H\leq7.25$ T ($H_c$ is the critical field where the moments are partially polarized), the $\lambda$-peak centered at $T_N$ gradually evolves into the hump-like behavior in $1/T_1(T)$, indicating suppression of the critical fluctuations by the applied magnetic field. For higher field intensities, the gapped spin excitation dominates the low temperature SLRRs as indicated by the thermally activated behavior. The gap size shows a linear field dependence up to the present 33 T with a negative intercept. Based on the two-dimensional spin coupling model, this indicates prominent single ion anisotropy with a positive value, demonstrating the easy-plane anisotropy of the spin system.

The polycrystalline sample of Na$_2$Cu$_3$O(SO$_4$)$_3$ was synthesized by a solid-state reaction method as described elsewhere\cite{Fujihala_PRL_120_077201}. Our NMR measurements were conducted on $^{23}$Na nuclei ($\gamma_n=11.262$ MHz/T, $I=3/2$) with a phase-coherent NMR spectrometer. The magnetic field below 16 T was generated by a superconducting magnet with high homogeneity, and a water-cooled resistive magnet was used for higher magnetic fields. The spectrum was obtained by integrating or summing up the spectral weight by sweeping the frequency or field with the other fixed. The SLRR ($1/T_1$) was measured with the inversion-recovery sequence, and calculated by fitting the time dependence of the nuclear magnetization ($M(t)$) to $M(t)/M(\infty)= 1-0.1\exp[-(t/T_1)^{\beta}]-0.9\exp[-(6t/T_1)^{\beta}]$. The stretching exponent $\beta$ reflects the distribution of SLRR in the sample.

\begin{figure}
\includegraphics[width=8cm, height=8cm]{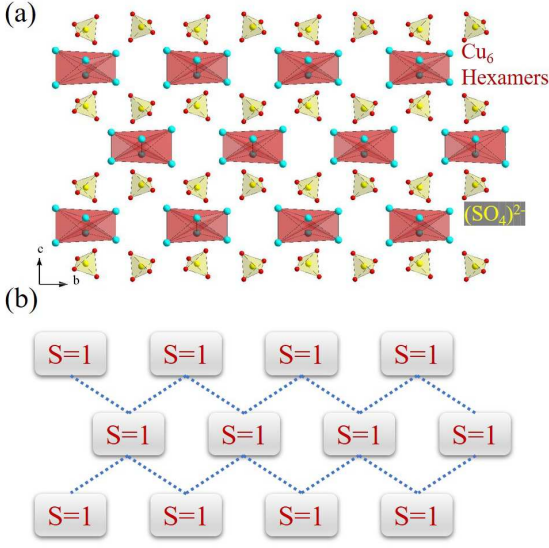}
\caption{\label{struc1}(color online) (a) Schematic crystal structure of Na$_2$Cu$_3$O(SO$_4$)$_3$ as seen against $a$-axis. The Na atoms are not
  shown for the clearance. (b) A sketched description of the two dimensional magnetic coupling between the spin-clusters carrying $S=1$.
}
\end{figure}

The Na$_2$Cu$_3$O(SO$_4$)$_3$ adopts a monoclinic structure with the $C2/C$ structure\cite{Fujihala_PRL_120_077201}. In Fig.\ref{struc1}(a), we show the crystal structure viewed against $a$-axis according to the atomic coordinates\cite{Nekrasova_PRB_102_184405}. Cu$_6$-hexamers are shown by the dark red polyhedra. The energy gap between the triplet ground state and first excited singlet state is larger than 12 meV by the recent INS results\cite{Furrer_PRB_98_180410}. As a result, the low temperature magnetism can be treated as interacting spin clusters ($S=1$) with high accuracy. By chelation with nonmagnetic oxyanions (SO$_4$)$^{2-}$, the inter-hexamer interactions are about two order of magnitude weaker than the intra-hexamer ones\cite{Furrer_PRB_104_L220401}. We schematically show the two-dimensional spin coupling between the nearest cluster neighbors in Fig.\ref{struc1}(b) as proposed by the calculations and INS measurements\cite{Nekrasova_PRB_102_184405,Furrer_PRB_104_L220401}.

\begin{figure}
\includegraphics[width=8cm, height=8cm]{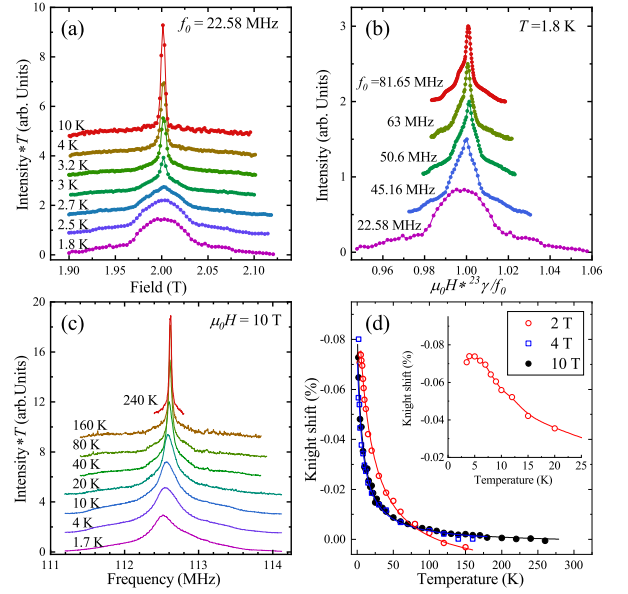}
\caption{\label{spec2}(color online)
  (a) Field-swept $^{23}$Na NMR spectra at different temperatures with the resonance frequency $f_0$ fixed at 22.58 MHz.
  (b) $^{23}$Na NMR spectra at $T=1.8$ K with different fixed frequencies $f_0$. The corresponding resonance fields without line shift are 2 T, 4 T, 4.5 T, 5.6 T and 7.25 T, respectively.
  (c) Typical $^{23}$Na NMR spectra under $\mu_0H=10$ T at various temperatures.
  (d) Temperature dependence of Knight shifts under different field intensities. Inset: A enlarged version of the Knight shifts at $\mu_0T=2$ T.
}
\end{figure}

In Fig.\ref{spec2}(a) and (c), we show $^{23}$Na NMR spectra at different temperatures respectively under $\mu_0H\sim2$ T and $\mu_0H=10$ T. The former are obtained by integrating the spectral intensity during sweeping field, while the later are obtained by summing up the spectral weight with the frequency swept. Each spectrum consists of a single central peak and a broad background. At $T=10$ K, the full width at half maximum (FWHM) of the central peak is $49.3$ Gauss ($\sim52$ kHz) for $\mu_0H\approx2$ T and 286 kHz for $\mu_0H=10$ T, nearly proportional to the field intensity. However, the spectral width of the background signal is $\sim2$ MHz for both fields, showing negligible field dependence. The nuclear quadrupole interaction with non-zero local electric field gradient (EFG) results in the satellite peaks, whose frequency distance from the central peak is very sensitive to the applied field direction, while independent on the field intensity in the strong field regime\cite{Abragam_book}. The broad background signal is the consequence of the random alignment between the field direction and the principle axis of EFG, and the quadrupole frequency $\nu_Q$ is roughly estimated to be $\sim2$ MHz\cite{SM}. The central peak line width is mainly contributed by anisotropic and distributed Knight shift (relative line shift with respect to the Larmor frequency).

The magnetic order under low magnetic fields can be detected by $^{23}$Na spectra. With sample cooling down, the central peak shows a sudden broadening below $T_N=3$ K, and further develops into a line shape with flattop at $T=1.8$ K (See Fig.\ref{spec2}(a)). This phenomenon indicates setup of the static internal field, also the occurrence of magnetic ordering\cite{Yamada_JPSJ_55_1751}. We further show the NMR spectrum at $T=1.8$ K under different magnetic fields in Fig.\ref{spec2}(b). To get rid of the central peak broadening effect from anisotropic Knight shifts, the field-swept spectra are shown with the field intensity divided by the resonance field of the fixed frequency. With the field increasing to 4 T, the flattop feature disappears, the sharp peak restores, and the relative line width gets narrower. When the field goes higher, the central peak becomes sharper and develops to be asymmetric. The evolution with field intensity is caused by the magnetic moment alignment by applied field and demonstrates a partially polarized state under moderate field intensities. This is consistent with the weak kink observed from the field dependence of magnetization\cite{Nekrasova_PRB_102_184405}. The asymmetric line shape in the paramagnetic state is the typical powder pattern resulting from anisotropic Knight shift\cite{Higa_PRB_96_024405}. We calculate the Knight shift at the most sharp part of the central line, and show its temperature dependence in Fig.\ref{spec2}(d). All the Knight shifts show an upturn behavior with temperature decreasing. The Knight shifts under $\mu_0H=2$ T show a maximum at $T\sim5$ K, consistent with the previous magnetic susceptibility measurements\cite{Nekrasova_PRB_102_184405}.

\begin{figure}
\includegraphics[width=8cm, height=6cm]{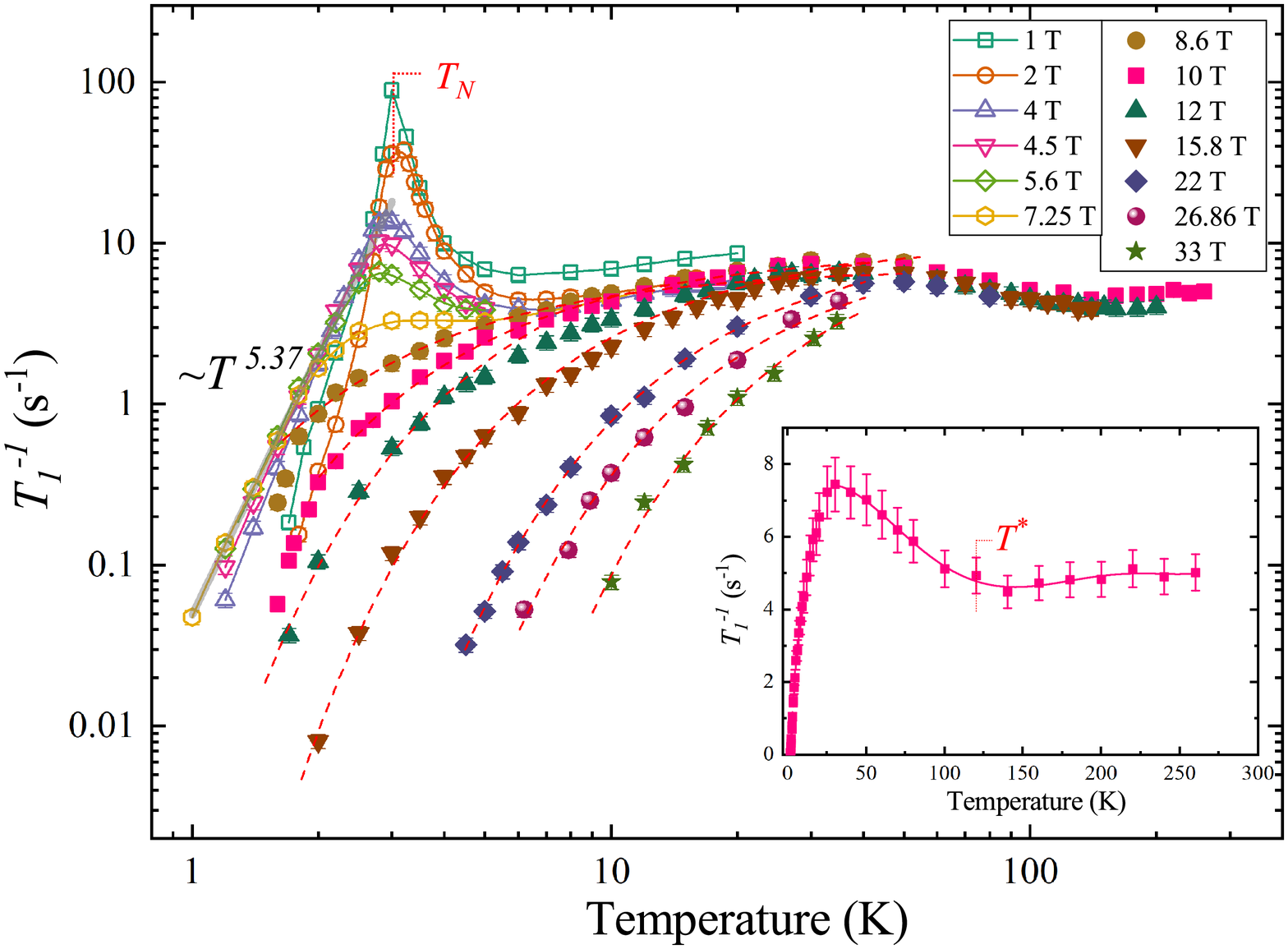}
\caption{\label{relax3}(color online)
  The temperature dependence of the spin-lattice relaxation rate $1/T_1$ for a wide magnetic field range. We use log-log plot here to show the drastic temperature dependence of $1/T_1$ at low temperatures more clearly. The gray straight line is the power law fitting to the low temperature $1/T_1$ for $\mu_0H=5.6$ T and 7.25 T. The linear plot of $1/T_1(T)$ at $\mu_0H=10$ T is shown in the inset to emphatically the overall $1/T_1(T)$ at high temperatures.
}
\end{figure}

Next, we begin to study the inter-hexamer magnetic interaction via the evolution of low-energy spin excitations with applied magnetic fields. In Fig.\ref{relax3}, we show $^{23}$Na SLRR $1/T_1$ as a function of temperature with magnetic fields ranging from 1 T
to 33 T. The temperature dependence of the stretching exponent $\beta$ and some typical fitting curves are shown in the supplemental materials\cite{SM}.
With the sample cooling down from $T=260$ K to 140 K, weak temperature dependence is observed from $1/T_1(T)$, which is followed by
an upturn behavior below the crossover temperature $T^*\sim120$ K (See Fig.\ref{relax3} inset). This crossover behavior is consistent with the slope change observed from the magnetic susceptibility measurements\cite{Nekrasova_PRB_102_184405}, both of which should result from frozen out of the excited singlet state of Cu$_6$-hexamer at low temperatures. With temperature far below $T^*$, the $S=1$ spin-cluster based description is very precise for the spin system in Na$_2$Cu$_3$O(SO$_4$)$_3$.

Prominent field dependence is observed from the $1/T_1(T)$ in the low temperature range. Under a $\mu_0H=1$ T or 2 T below the critical field $H_c$\cite{Nekrasova_PRB_102_184405}, the $1/T_1$ shows rapid rise below $T=5$ K, then further drops steeply below $T_N$, forming a sharp peak centered at $T_N$. This behavior results from the critical slowing down of spin fluctuations near the phase transition\cite{Moriya_PTP_28_371,Ziolo_JAP_63_3095}, widely observed in various magnetic compounds. With the applied field intensity exceeding $H_c$(e.g. $\mu_0H=4$ T), a hump behavior still exist as a shadow of previous sharp peaks in the low temperature $1/T_1(T)$, which survives to a higher field $\mu_0H=7.25$ T. This phenomenon reflects the suppression of the critical spin fluctuations by applied magnetic field.

The low temperature $1/T_1$ at $\mu_0H=5.6$ T and 7.25 T precisely shows a power-law temperature dependence as shown by the straight line in Fig.\ref{relax3}. The scattering of magnons off nuclear spins dominates the low temperature $1/T_1$ in this magnetic insulator. For $T\gg\Delta$ where $\Delta$ is the magnon gap, $1/T_1$ follows either a $1/T_1\propto T^3$ behavior due to a two-magnon Raman process or a $1/T_1\propto T^5$ behavior due to a three-magnon process\cite{Beeman_PR_1968}. For $T\ll\Delta$, the $ 1/T_1$ will show a thermally activated behavior. The persistent power-law behavior of $1/T_1(T)$ down to $T=1$ K suggests a very small magnon excitation gap, which is consistent with the field dependence of the magnon gap as discussed below. The power law index of 5.37 suggests that the nuclear relaxation is dominated by three-magnon Raman process.

\begin{figure}
\includegraphics[width=8cm, height=6cm]{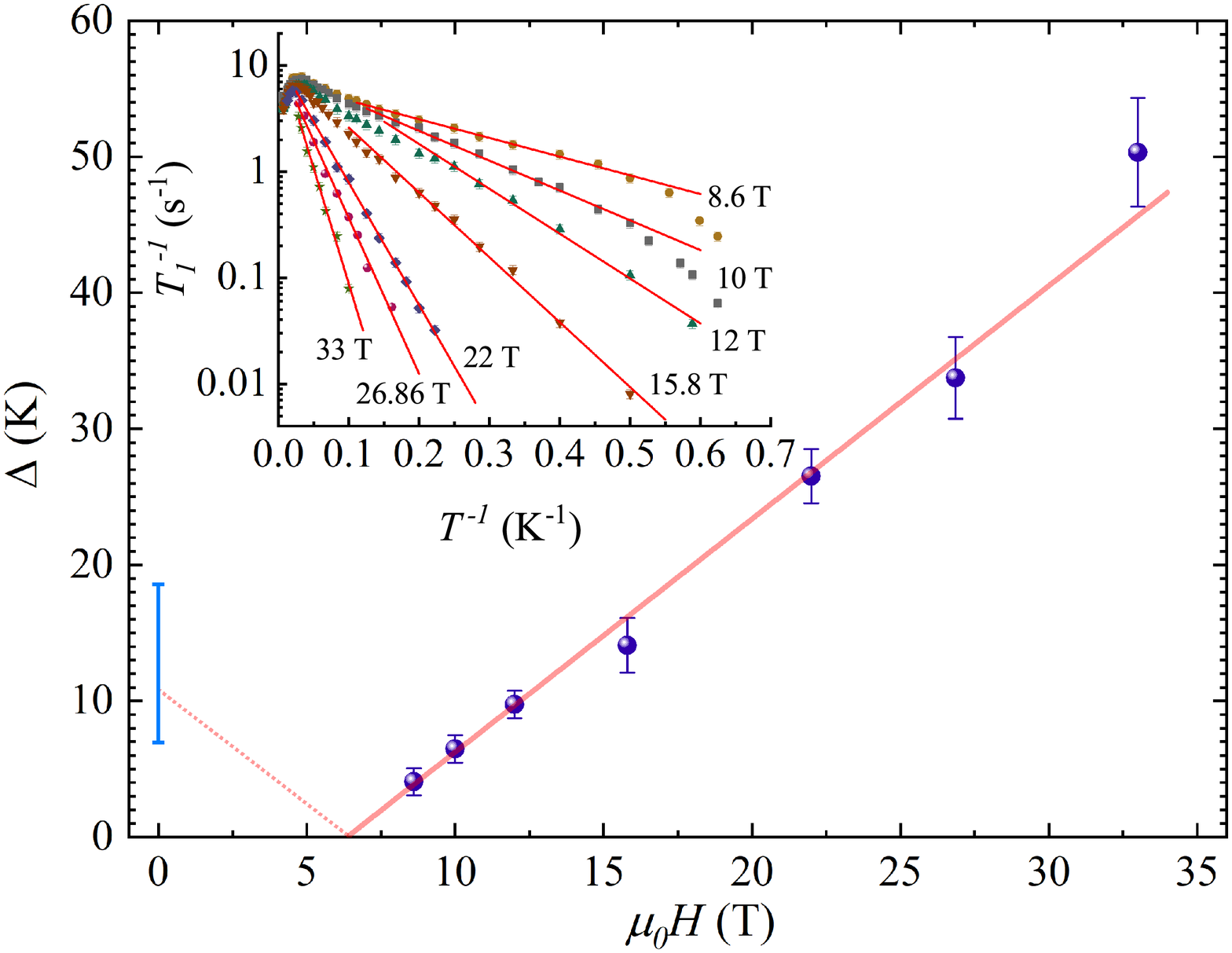}
\caption{\label{gap4}(color online)
  The spin excitation gap size (in K) as a function of the applied magnetic field intensity. The solid straight line is a linear fit to the data.
  Inset: The $1/T_1$ plotted as a function of the inverse temperature $1/T$ on semilog coordinate. The red lines show the gapped behavior shown by $1/T_1(T)$. The zero field spin gap size range given by INS measurements is marked by the light blue bar at the left side.
}
\end{figure}

Thermally activated temperature dependence of $1/T_1$ is observed after the critical fluctuation is fully suppressed by applied magnetic fields.
The $1/T_1(T)$ can be well fitted by the gapped function $1/T_1=A\exp(-\Delta/T)$, which is demonstrated by the dashed lines in Fig.\ref{relax3}.
To show this gapped behavior more clearly, we replot the $1/T_1$ versus the reciprocal of temperature ($1/T$) in a log-linear scale (See Fig.\ref{gap4} inset). The slope of straight lines is directly proportional to the gap size $\Delta$, whose field dependence is shown in Fig.\ref{gap4}.
The field dependence of the gap size can be well reproduced by a simple linear function, which is shown by the solid line. The slope and intercept are determined to be 1.71 K/T and -10.86 K, respectively. The continuous linear dependence for $8.6$ T $\leq\mu_0H\leq33$ T indicates the $S=1$ spin-cluster description holds true for the studied field range.

The field dependence of the spin excitation gap can be understood with the Heisenberg spin model with single-ion anisotropy. The Hamiltonian can be written as,
\[
\begin{split}
H=D\sum_i(S_i^z)^2+g\mu_BB\sum_iS_i^z+\\
J\sum_{i,j}[S_i^zS_j^z+\frac{1}{2}(S_i^+S_j^-+S_i^-S_j^+)].
\end{split}
\]
The $D$ parameterizes the single-ion anisotropy, determining the easy-axis or easy-plane spin orientation. The $J>0$ measures the antiferromagnetic exchange coupling strength, which is treated as isotropic firstly for simplicity. The second term describes the Zeeman splitting raised by applied field.

For $D\gg nJ$ ($n$ is the coordination number), the $S^z=0$ state is energetically favorable. The spin excitation energy $\Delta E(\mathbf{k})$ of the collective spin wave can be expressed as $\Delta E(\mathbf{k})=D+nJ\gamma_{\mathbf{k}}\mp g\mu_BB$, where $\gamma_{\mathbf{k}}$ denotes the structure factor\cite{Ghosh_PRB_88_094414}. With the anisotropic exchange coupling strength $J$ included, similar field dependence of the spin gap size can be also obtained just with a modified spin gap at zero field\cite{Ghosh_PRB_90_224405}. The negative branch fits quite well with the field dependence of the spin excitation gap show in Fig.\ref{gap4}.
The slope of the solid straight line yields a magnetic moment of 2.54 $\mu_B$, comparable with the reported 2.84 $\mu_B$ determined by low-temperature magnetic susceptibility. By reversing the fitting line to the positive $\Delta$-axis symmetrically, the zero field gap $\Delta_0$ is suggested to be 10.86 K (0.94 meV), just locating in the gap range $0.6\sim1.6$ meV observed from INS\cite{Furrer_PRB_104_L220401}.

One essential basis for the above analysis is the existence of collective magnons or paramagnons. In conventional three-dimensional magnets, the spin wave excitation gap in the long-range ordered state originates from the spin anisotropy, which closes above $T_N$ due to thermal fluctuations\cite{Nekrasova_PRB_102_184405}. In the present Na$_2$Cu$_3$O(SO$_4$)$_3$, the zero field spin gap at $T=1.5$ K below $T_N$ can be described as a consequence of the Heisenberg spin model with single-ion anisotropy\cite{Nekrasova_PRB_102_184405}. With the temperature exceeding $T_N$, although the spin excitation peak is broadened and in-gap states appear, the overall characteristics for the spin gap are maintained in the energy spectra of neutrons\cite{Furrer_PRB_98_180410}. This suggests the existence of short range order above the ordering temperature. For $\mu_0H>8.6$ T, the spin gap exceeds the energy scale defined by $T_N$, and finally reaches $\Delta\sim50$ K at $\mu_0H=33$ T. Based on the verified two-dimensional magnetic interaction in the present sample, we propose the persistent short range order under applied magnetic fields, which supports the paramagnon excitations above $T_N$.

Short range order in magnetic materials usually result from magnetic frustration or/and anisotropic spin coupling. In the classical frustrated antiferromagnet YMnO$_3$ with an ordered ground state, persistent paramagnon excitations are observed far above the Neel temperature in a short-range manner\cite{Demmel_PRB_76_212402}. In rare-earth kagome systems, Nd$_3$Ga$_5$SiO$_{14}$, Pr$_3$Ga$_5$SiO$_{14}$ and Pr$_3$BWO$_9$ with cooperative paramagnetic ground state, the paramagnon excitations in the dynamical spin-clusters are verified by magnetic resonance studies\cite{Ghosh_PRB_90_224405,Ghosh_PRB_88_094414,Zeng_PRB_104_155150}. For anisotropic coupled spin chain CsMnCl$_3$, prominent spin excitations are found to persist to $2T_N$\cite{Skalyo_PRB_2_4632}. Even in the two-dimensional ferromagnet CrI$_3$, the paramagnon excitations also appear above $T_C$\cite{Chen_PRB_101_134418}, where these Goldstone modes are more unstable than that in antiferromagnets. In the present Na$_2$Cu$_3$O(SO$_4$)$_3$, anisotropic magnetic interactions are indicated by the existing spin excitation gap and its field dependence. The short-range order should mainly be caused by the anisotropic spin coupling. The short range order persists to higher temperatures under stronger magnetic fields, which is consistent with the increasing spin gap. The magnetic frustration is likely absent here as indicated by the previous calculations and INS results\cite{Nekrasova_PRB_102_184405,Furrer_PRB_104_L220401}.

Finally, we discuss connections between the zero field spin gap and the present high field one. From our study, the spin excitation gap at zero field
is predicated by the linear fitting intercept, exactly matching the gap size reported by INS study\cite{Furrer_PRB_104_L220401}. This fact suggests common magnetic interactions for the entire magnetic field range, together emphasizing the spin anisotropy of the Cu$_6$ spin-clusters.

In conclusion, we have performed NMR study on the spin-cluster based mineral Na$_2$Cu$_3$O(SO$_4$)$_3$ within a ultra-wide magnetic field range. Under low fields, the long-range magnetic order is monitored by a sudden broadening of the resonance line across $T_N$. Correspondingly, the critical slowing down behavior is observed near $T_N$ from $1/T_1(T)$. The hump behavior in $1/T_1(T)$ persists to $\mu_0H=7.25$ T, much higher than $H_c$ where the magnetic moments are partially polarized. With higher field up to the studied $\mu_0H=33$ T, the low temperature $1/T_1(T)$ is dominated by a thermally activated behavior, indicating gapped spin excitations in the high field region. The gap size show a linear field dependence, whose slope and intercept respectively yields an effective magnetic moment of 2.54 $\mu_B$ and 0.94 meV spin excitation gap at zero field. The high field spin gap and its field dependence indicate the underlying paramagnon excitations supported by short range order and prominent easy-plane spin anisotropy which should play an important role in the spin excitation spectrum of A$_2$Cu$_3$O(SO$_4$)$_3$.

This research was supported by the National Natural Science Foundation of China (Grants No. 11874057, 21927814 and 21972145) and the Collaborative Innovation Program of Hefei Science Center, CAS (2021HSC-CIP002). A portion of this work was supported by the High Magnetic Field Laboratory of Anhui Province.


\end{document}